\documentclass[11pt]{article}
\usepackage[a4paper, top=2.0cm, bottom=2.0cm, left=2.0cm, right=2.0cm]{geometry}
\usepackage{multicol}
\usepackage{setspace}
\usepackage{graphicx}
\usepackage{amsmath,amssymb}
\usepackage{wrapfig}
\usepackage{chemformula}
\usepackage{caption}
\usepackage{upgreek}

\usepackage{siunitx}

\usepackage{enumitem} 
\usepackage{paralist}
\setitemize{noitemsep,topsep=0pt,parsep=0pt,partopsep=0pt,leftmargin=1.5\parindent}
\usepackage{hyperref}
\definecolor{darkorange}{RGB}{230, 120, 0}
\hypersetup{
    colorlinks=true,
    linkcolor=darkorange,
    urlcolor=darkorange,
    citecolor=darkorange,
}
\usepackage{natbib}
\usepackage{aasmacros}
\usepackage{doi}

\usepackage{soul}

\newcommand{\squishlist}{
 \begin{list}{$\bullet$}
  { \setlength{\itemsep}{1pt}
     \setlength{\parsep}{0pt}
     \setlength{\topsep}{1pt}
     \setlength{\partopsep}{0pt}
     \setlength{\leftmargin}{1.5em}
     \setlength{\labelwidth}{1.5em}
     \setlength{\labelsep}{0.5em} } }
\newcommand{\squishend}{
  \end{list}  }

\begin{document} 

\title{The power of polarimetry for characterising exoplanet atmospheres, clouds, and surfaces with NASA's Habitable Worlds Observatory (thematic area: Astro)} 
\author{Katy L. Chubb (University of Bristol)\footnote{katy.chubb@bristol.ac.uk},\\ Mei Ting Mak (University of Oxford, University of Exeter),\\ Joanna K. Barstow (The Open University), Beth Biller (University of Edinburgh),\\ Sarah Rugheimer (University of Edinburgh), Daphne M. Stam (Leiden Observatory),\\ Victor Trees (Delft University of Technology)}

\date{}

\maketitle

\vspace{-4mm} 

\section{Scientific Motivation \& Objectives}

Building on the recent success of JWST, and the upcoming Ariel mission, in which the UK has a strong presence in both the instrument and science development, the Habitable Worlds Observatory (HWO) will be the next breakthrough in exoplanet atmosphere characterisation. Due for launch in the 2040s, HWO will allow detailed observation of Earth-like exoplanet atmospheres and surfaces for the first time~\footnote{\url{https://science.nasa.gov/astrophysics/programs/habitable-worlds-observatory/}}. 
We are now at a critical stage of being able to inform the instrument and mission design of HWO. There are three principal instruments very likely to be included on HWO, including a coronagraph for high-contrast imaging and imaging spectroscopy, and a high-resolution imager. The coronagraph instrument is required for observing Earth-like exoplanets due to their relatively faint reflected flux signal, whereas larger exoplanets orbiting closer to their host star will be observed by an instrument such as the HRI. Although the coronagraph will be NASA-led~\footnote{\url{https://www.jpl.nasa.gov/habex/mission/instruments/coronagraph/}}, there is potential for international collaboration and it will be possible for UK scientists to contribute to its hardware and design, in particular the infrared channel. The UK Astronomy Technology Centre (UK ATC) is currently leading a design study for an infrared spectrograph for the coronagraph IR arm.
The teams that will lead and contribute to the design and hardware of the high-resolution imager (HRI) have not been set. Currently, two UK-based teams are leading independent studies funded by UKSA with the aim of securing UK leadership of the HRI for HWO. Whether a polarimeter is included on these instruments is still to be decided.
Other proposals include a high resolution spectra-polarimeter, Pollux, which is a French-led proposed instrument concept, with R$>$60,000 and covering $\lambda$~$\sim$0.12-1.8~$\mu$m. The Pollux team and working groups include UK-based scientists, on both the instrument and the modeling side.  

\begin{figure*}
\centering
\includegraphics[width=0.7\textwidth]{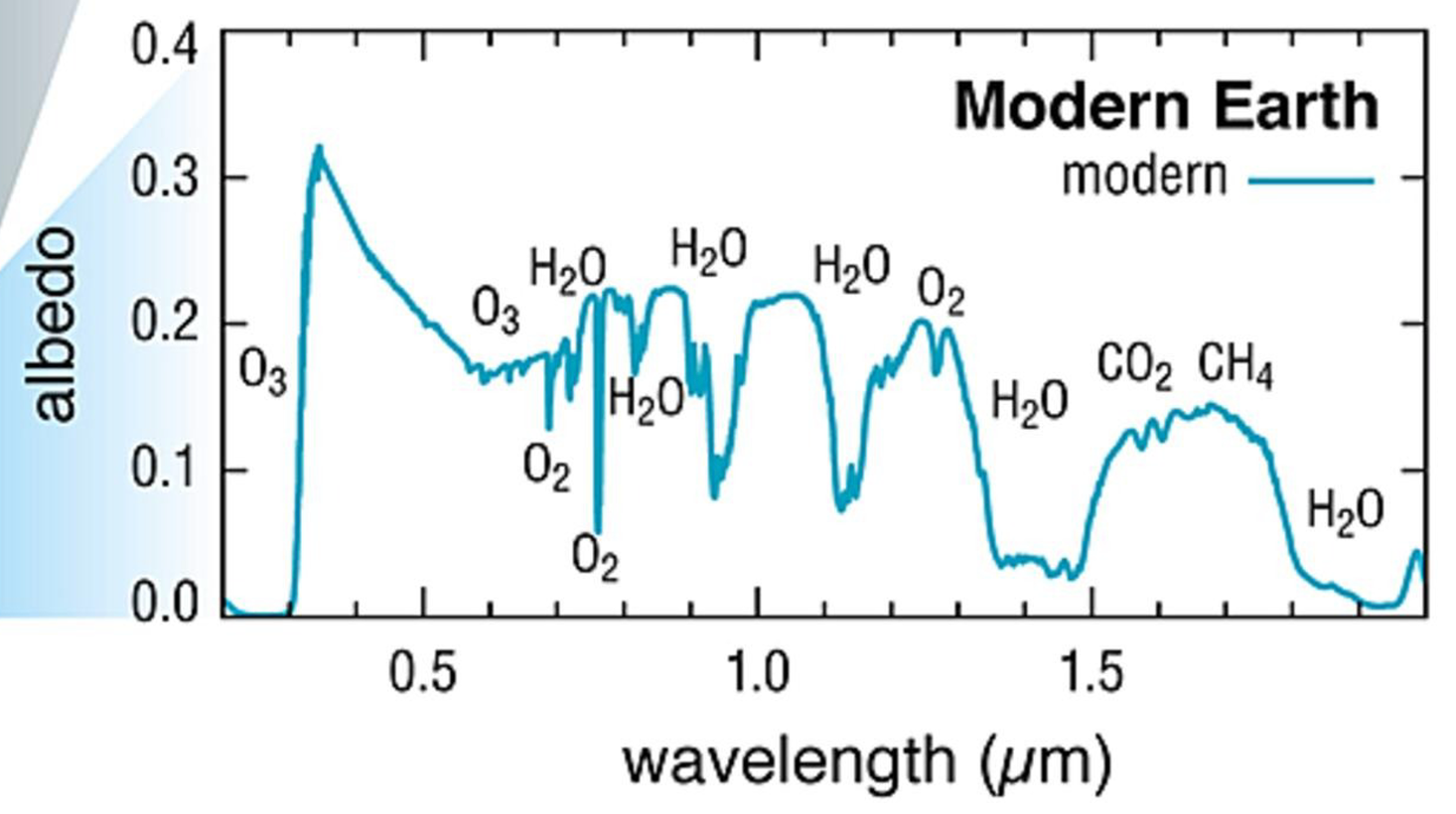}
\caption{The reflectance spectrum detailing features from O$_2$, O$_3$, H$_2$O, CO$_2$, and CH$_4$ of a modern Earth-like planet from 0.2~-~2~$\mu$m, similar in range expected from HWO (adapted from \cite{Westall2024}).}
\label{fig:earth_albedo}
\end{figure*}

HWO will be a powerful tool for understanding exoplanet atmospheres and surfaces, via reflection spectroscopy. Reflection spectroscopy is a measure of the light which is reflected from a star by a planet's atmosphere or surface towards the observer, as a function of wavelength. The refractive properties of the material, either clouds or gases in the atmosphere, or the planet's surface (which can include liquid water), determine the reflection as a function of wavelength~\citep{Kawashima_2019}. Analysis of the reflected light from an exoplanet, ideally as it rotates or moves through its orbit and thus presents different viewing angles, tells us about the properties of the atmosphere and, if applicable, surface of the planet. As can be seen by Figure~\ref{fig:earth_albedo}, which is a model of the albedo (reflected light) spectrum of modern Earth, the fact that HWO covers the UV and optical region, a regime that is not achievable with high sensitivity by any current space telescopes, allows us access to observe absorption features from molecules such as O$_2$, O$_3$, H$_2$O, CO$_2$, and CH$_4$, essential for assessing a planet's habitability. Other interesting aspects of habitability, including the vegetation red-edge, can also be accessed via reflection spectroscopy~\citep{18MaKa}.
Clouds are an essential component of any planet's atmosphere. They dictate a planet's energy budget and therefore have a huge impact on the planetary climate. We will not be able to assess a planet's habitability without properly diagnosing the cloud type, size, location and spread throughout the atmosphere~\citep{Yang_2025}. It is therefore crucial that we are using the right instruments for observing and theoretical tools for analysing the atmospheres that allow us to extract these cloud properties.

\begin{figure*}
\centering
\includegraphics[width=0.8\textwidth]{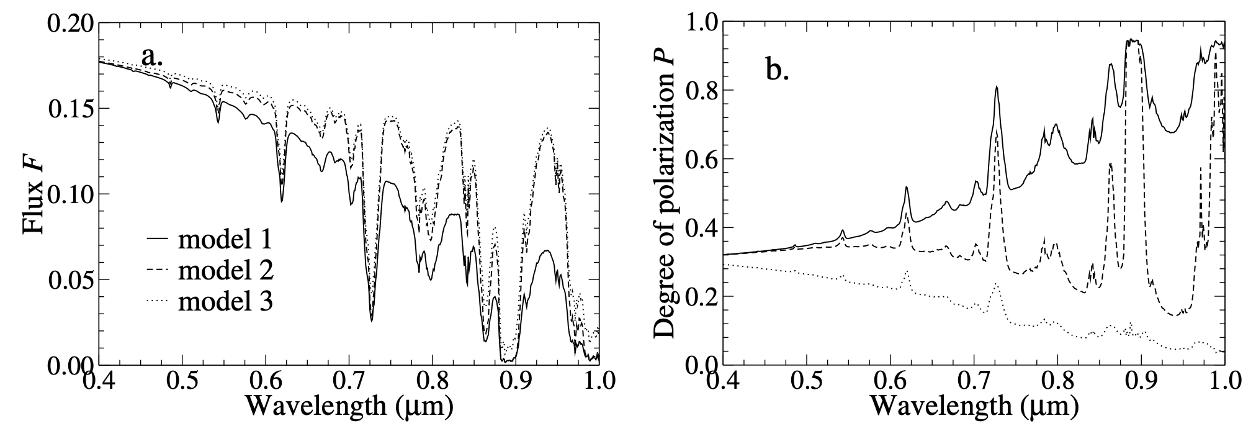}
\caption{The reflected flux (left) and degree of polarisation (right) (i.e. proportion of the flux that is linearly polarised at a given wavelength) of a Jupiter-like exoplanet for three model atmospheres at an orbital phase of 90$^{\circ}$. Model 1: gas only (no clouds), model 2: the same as model 1 plus a tropospheric cloud layer, model 3: the same as model 2 plus a stratospheric haze layer (from \cite{04Stam}).}
\label{fig:jupiter}
\end{figure*}

\begin{figure*}
\centering
\includegraphics[width=1.0\textwidth]{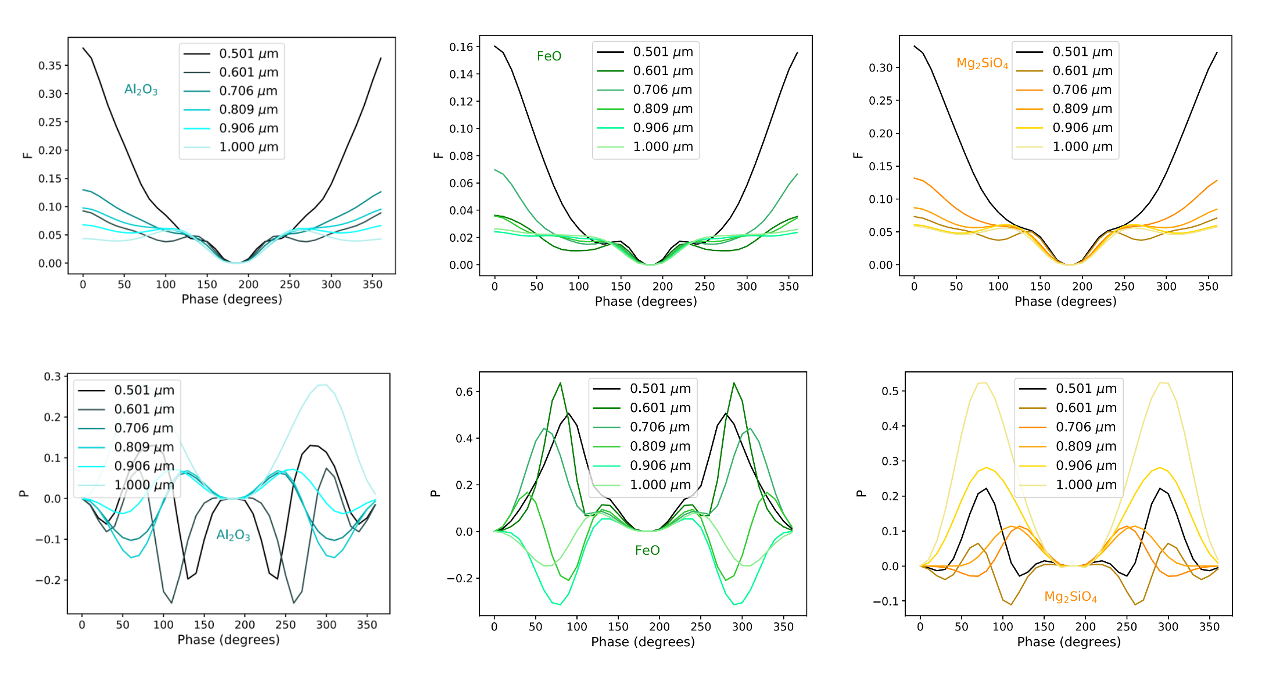}
\caption{The reflected flux (upper panels) and degree of polarisation (lower panels) as a function of orbital phase of a Hot-Jupiter exoplanet for model atmospheres containing clouds made up of different materials: Al$_2$O$_3$ (right), FeO (middle), and Mg$_2$SiO$_4$ (right). Different wavelengths are shown for each model. This figure is adapted from \cite{23ChStHe}.}
\label{fig:polhex}
\end{figure*}

\textbf{Spectropolarimetry} is an extra measure of the polarisation state of the reflected light as a function of wavelength. The polarisation state describes the direction in which the light’s electric field oscillates as the wave propagates. In the case of linear polarisation, the electric field oscillates preferentially along a single plane. By measuring how the degree of linear polarisation varies with wavelength and planetary phase angle, we gain substantially stronger constraints on key exoplanet properties, including cloud composition, particle size and shape, and surface type~\citep{74HaTr}.
For example, Figure~\ref{fig:jupiter} shows the total flux (left) and the degree of linear polarisation (i.e. proportion of the flux that is linearly polarised at a given wavelength) (right) of light that is reflected by 3 model Jupiter-like exoplanets. 
It can be seen that models 2 and 3 in particular are very hard to differentiate if only the total (unpolarised) flux were measured, but if the polarisation state is also collected then the spectra in the right panel show these models become more easily distinguishable, provided the instrument has a high-enough accuracy in the polarisation measurement. Figure~\ref{fig:polhex} shows the reflected flux (upper panels) and degree of polarisation (lower panels) for a modelled close-in hot-Jupiter exoplanet as a function of orbital phase and for various wavelengths between 0.5~-~1~$\mu$m for different cloud species. It can be seen that it is hard to distinguish between the different cloud species using reflected flux alone in this wavelength region, but using the degree of polarisation can really help break these degeneracies between models.

Even without polarimetry included in one of its instruments, HWO will allow a wealth of information on exoplanets, particularly for smaller, Earth-like exoplanets currently inaccessible by other instruments. However, including the polarisation state of the reflected light unlocks a whole new level of detail on the atmosphere, clouds, and surface of the exoplanet. 
\textbf{Collecting the polarisation state (i.e. the proportion of the flux which is linearly polarised as a function of wavelength) of reflected spectra gives all the information that would be gained from reflection spectra, with additional information content (i.e. cloud particle size, shape, composition, and vertical distribution, as well as surface type) to be analysed and interpreted.}

A number of studies have demonstrated the diagnostic power of polarised reflection spectra for characterising a wide range of exoplanet properties. For potentially habitable planets, polarimetry can be used to identify surface features such as ocean glint~\citep{19TrSt,22TrSt,23VaGeBo}, to probe atmospheric composition through features such as the O$_2$ A-band~\citep{17FaRoSt}, and to constrain the properties of liquid water clouds~\citep{11KaStHo}. It has also been applied to modelling Earth-like exoplanets across different stages of their evolutionary history~\citep{08Stam,Goodis_Gordon_2025}.
The benefits of polarimetry for more detailed characterisation of hotter, larger, planets orbiting closer to their host star have also been demonstrated~\citep{23ChStHe}; and for the observation of solar system analogue planets such as Venus~\citep{23MaAbRo}. Observing and analysing the reflection spectra, and in particular the polarised reflection spectra, is a complementary technique to the current observations and analyses of transiting exoplanets with JWST, Ariel, or the Hubble Space Telescope (HST). 

The HWO addresses some high-priority scientific questions that the UK should pursue in the next decade, including ``how common are the planets in our solar system, in particular Earth, compared to other planets in our galaxy''. We, as exoplanet scientists in the UK, have a unique opportunity to be involved in answering this question.

\section{Strategic Context}

A number of Science Case Development Documents (SCDDs) were recently written for HWO, including some significant UK-led, or with strong UK-contributions, such as \cite{wakeford2025_SCDD}, signifying that the UK is already well represented in the science development. There is an active community networking for HWO, bringing instrument and science developers together. A number of these SCDDs mention polarimetry as a valuable and desirable tool to include on at least one of the HWO instruments~\footnote{\href{https://docs.google.com/spreadsheets/d/1PTazkPP-gIhOEETNVDLoXp-7m-1etTRdknWWlmQrPNI/edit?usp=sharing}{SCDD Portal Table}}.

There are already, and will continue to be, opportunities for partnerships between academia and industry, particularly for continuing to build a case for a UK-led, or a strong UK-contribution, to a proposed instrument for HWO. Apart from the US-led coronagraph, which has potential for UK contribution, the other instruments are still in the planning and proposal stage. This is expected to include a high resolution imager (HRI), and a UV multi-object spectrograph (MOS), which have the potential to be a UK-led instruments both in terms of the technology and science expertise.  
The HWO is due for launch in the 2040s, placing us now in a crucial time for technological development. The Mission Concept Review (MCR) is due in 2029, a deadline by which all crucial technologies need to be at a Technological Readiness Level (TRL) of at least 5. This requires the first-generation instruments that will be included in the mission to be confirmed by 2029. 
The UK being a core part of the first mission to observe potentially habitable Earth-like exoplanets will be a huge inspiration and driver for bringing the next generation of students into the space industry, and encouraging technological advancement that places the UK in a world-leading position. 

\section{Proposed Approach}

Different types of exoplanets will be observed by different instruments onboard HWO. The coronagraph is required for observing Earth-like exoplanets. The strength of the reflected light signal increases with planet size, star size and brightness, and decreases with increasing planet-star distance. The signal is thus fainter for smaller planets orbiting further out from their host star, i.e. Earth-like planets. These planets require the use of the coronagraph to block out the stellar light in order to observe the relatively faint reflective light. The closer a planet is to the host star the stronger the reflected signal becomes, with larger planets also having a stronger signal. However, this means the planets are not sufficiently separated from the host star to be observed using a coronagraph. Fortunately the stronger signal of closer, larger planets means they do not require a coronagraph, and can be observed with an instrument
such as the proposed High Resolution Imager (HRI) with a grism which allows for spectroscopic observation. Another proposed instrument, Pollux, is a candidate French-led HWO instrument proposed to have a polarimeter, which currently has contributions from a variety of European countries including the UK. 
Pollux’s specifications can still be adjusted in response to its science objectives, as long as they are technically achievable. Including the option to include or omit a coronagraph would allow it to observe the broadest possible range of exoplanets.
By observing the phase curve with these instruments, where the combined reflected light of the planet and star's light are observed as a function of planet's orbit and wavelength, the planet's reflected light can be disentangled from the stellar light. The planet's reflection spectra can therefore be extracted for analysis. 

Polarisation is a relative measure, so as long as there is enough flux reflected by the exoplanet towards us, and assuming the star can be considered unpolarised when integrated over the stellar disk (i.e. assuming the star is inactive~\citep{87KeHeSt,17CoMaBa}), then we should be able to detect relatively small proportions of polarised flux. An instrument capable of polarimetry or spectropolarimetry should have the ability to detect a linearly polarised signal of ideally at least 0.1~--~1ppm (as a proportion of total flux) to be able to observe a range of exoplanets.
As a comparison, HIPPI-2, which is a ground-based telescope which includes a polarimeter, has a precision of 3.5ppm~\citep{20BaCoKe}. 

In summary, we would target cooler, smaller, Earth-like planets using the coronagraph. We would target larger or closer-in planets which are either transiting or close-to-transiting with instruments likely to be included in the HWO such as HRI or Pollux, enabling the phase curve (monitoring the planet's spectra as a function of its orbit round its host star) to be used a strong indicator of the planet's atmosphere, including signatures of clouds and gases, and, if applicable, the planet's surface. To inform the design of these instruments, specific and accurate modelled reflection spectra which include polarisation are required for a range of exoplanets. 
This white paper advocates for a strong UK-presence in the science modeling and instrument design of Pollux, the HRI, and the Coronagraph.

\section{Proposed Technical Solution and Required Development}

As aforementioned, the Mission Concept Review is due in 2029, a deadline by which all crucial technologies need to be at a Technological Readiness Level (TRL) of at least 5. It is essential that we use this time to leverage the UK's rich expertise for influencing and contributing to (ideally leading) instrument design. This requires complementary development of theoretical models to back-up the design of these instruments, and to prove the ability to analyse the observed data well in advance of the MCR. 
Indirectly, the development of laboratory and theoretical data required for a mission such as HWO is crucial for the UK to continue to be at the forefront of exoplanet science. This includes high temperature and accurate molecular line lists~\citep{ExoMol2020}, and refractive indices for various cloud types. The UK-led Ariel space mission has a data working group which has a recent white paper~\citep{24ChRoSo} detailing the currently available and required data products for Ariel or a similar exoplanet characterisation mission like HWO. High performance computing (HPC) is another essential component of exoplanet science development, now and in the future, to process large amounts of data required for complex atmospheric models and observational analysis.

\section{UK Leadership and Capability}
The UK exoplanet community is very active. The UK has a lot of heritage in exoplanet science, with Plato and Ariel prominent UK-led ESA exoplanet missions. The UK is heavily involved in JWST, from leading the design and building of the MIRI instrument~\citep{Wright_2023} to UK-based scientists leading a number of important analyses (e.g. \cite{Alderson2023_ERS} and \cite{Ahrer2023_ERS} via the Early Release Science collaboration). In order to keep the UK at the forefront of exoplanet science and technology, an active involvement in HWO is essential. Various networks and meetings are already active which foster a strong collaborative environment, such as the UK exoplanet and HWO community meetings.  We will leverage this community to stay active and push forwards with HWO development, which is a key avenue for UK exoplanet scientists. 

\section{Partnership Opportunities}
Working on the HWO brings important opportunities for collaborations with scientists working in the US, in particular NASA, and a host of European countries. 
In addition, there are opportunities for collaboration with teams working on using polarisation to characterise planets in our solar system. For example, VenSpec-H is a spectrometer suite on board the ESA EnVision mission to Venus (due to launch in 2031)~\citep{VenSpecH}, which is expected to include polarimeters. The development of VenSpec-H is Belgian-led with opportunities for collaboration with other ESA State members. 

\newpage
\section*{Co-signers}

Lili Alderson (Cornell University, University of Bristol) \\ 
Amirnezam Amiri (University of Arkansas)  \\   
Jayne Birkby (University of Oxford) \\   
Marrick Braam (University of Bern) \\   
David J. A.  Brown (University of Warwick) \\   
Heather Cegla (University of Warwick) \\   
Trent Dupuy (University of Edinburgh) \\   
Charlotte Fairman (University of Bristol) \\
Carole Haswell (The Open University) \\   
Aiza Kenzhebekova (University of Edinburgh) \\   
Thaddeus Komacek (University of Oxford) \\      
Subhanjoy Mohanty (Imperial College London) \\   
Annelies Mortier (University of Birmingham) \\  
Mia Belle Parkinson (University of Edinburgh) \\        
Séverine Robert (Royal Belgian Institute for Space Aeronomy) \\    
Dominic Samra (University of Chicago) \\   
Jesper Skottfelt (The Open University) \\  
Clara Sousa-Silva (Bard College) \\
Jonathan Tennyson (University College London) \\
Ben Sutlieff (University of Edinburgh) \\       
Eleni Tsiakaliari (The Open University) \\   
Daniel Valentine (University of Bristol) \\     
Hannah Wakeford (University of Bristol) \\      
Sergey Yurchenko (University College London) \\

\bibliographystyle{aasjournal}
\bibliography{main}

\end{document}